\documentstyle[psfig]{acs}
\newcommand{\req}[1]{(\ref{#1})}
\newcommand{\be}{\begin{equation}}
\newcommand{\ee}{\end{equation}}
\newcommand{\bea}{\begin{eqnarray}}
\newcommand{\eea}{\end{eqnarray}}
\newcommand{\avg}[1]{\langle{#1}\rangle}
\newcommand{\ovl}{\overline}
\newcommand{\sign}{\textrm{sign}\ }

\begin{document}

\title[Volatility in toy markets]
{Trading behavior and excess volatility in toy markets}

\author[M. Marsili and D. Challet]{Matteo Marsili
\\
Istituto Nazionale per la Fisica della Materia (INFM)\\
Trieste-SISSA Unit,\\
V. Beirut 2-4, Trieste I-34014\\
\And
Damien Challet\\
Institut de Physique Th\'eorique, Universit\'e de Fribourg,\\
Perolles, CH-1700
}

\date{\date}
\pagerange{\pageref{firstpage}--\pageref{lastpage}}

\def\LaTeX{L\kern-.25em\raise.425ex\hbox{a}\kern-.075em\TeX}
\def\eg{{\rm e.g.\ }}
\def\etc{{\rm etc}}

\label{firstpage}
\maketitle

\begin{abstract}
{\em ABSTRACT\/}. We study the relation between the trading behavior
of agents and volatility in toy markets of adaptive inductively rational
agents. We show that excess volatility, in such simplified markets, 
arises as a consequence of {\em i)}
the neglect of market impact implicit in price taking behavior and of 
{\em ii)} excessive reactivity of agents. These issues are dealt with in 
detail in the simple case without public information. We also derive,
for the general case, the critical learning rate above which trading 
behavior leads to turbulent dynamics of the market.

\end{abstract}

\section{Introduction}

\noindent The El Farol bar problem \cite{Arthur} and its later refinement, 
the Minority Game (MG) \cite{CZ1}, 
where devised to study the competitive interaction of boundedly rational
agents in socio-economic contexts. 
In the El Farol bar problem agents face the 
binary choice of either attending or not a bar which is enjoyable only if
it is not too crowded. \ccite{Arthur} showed that 
inductive rationality is enough for the average attendance to 
self-organize around the comfort capacity of the bar. 

The El Farol bar context is simplified even further in the Minority Game.
Again agents compete in forecasting what the majority will do, in order
to be on the opposite side, which is the rewarding one. 
Agents,
like speculators in a financial market \cite{ZENews},
try to exploit {\em predictability} of the ``market''. This 
speculative pressure drives the system in a self-organized state
where, {\em on average}, half of the agents take each of the
two possible actions at each time. 

This confirms the naive expectation that speculative trading
eliminates {\em arbitrages} -- i.e. predictable patterns -- in
markets. In addition, the MG allows to focus attention on the fluctuations 
around the average and allows to address the question of whether and
to which extent adaptive agents reduce or amplify fluctuations. 
Fluctuations in the MG are directly related to the
profit losses and hence to wastes of resources. 
This is reminiscent of {\em volatility} in financial markets,
which is an undesirable consequence of speculative trading. 

The MG describes in a nutshell the interplay 
between predictability and volatility in a market with 
speculative trading. Since it allows to quantify both predictability 
$H$ and volatility $\sigma^2$ (see below), the MG allows to address 
questions of both information efficiency (predictability) and market 
stochasticity (volatility). 

Much numerical work \cite{CZ1,Savit,Johnson98,Cavagna,CM,Cavagna2} as well as 
approximate analytical approaches \cite{ZENews,CZ2,Johnson99} 
have been devoted to the understanding of 
the MG\footnote{A collection of references on the Minority Game is 
available at the site \tt{http://www.unifr.ch/econophysics/minority}.}. 
This has culminated in the exact solution \cite{CMZ,MCZ} in the limit of 
infinitely many agents, thanks to statistical mechanics tools of 
disordered systems \cite{MPV}. The emerging picture is that the speculative
trading of agents actually minimizes market's predictability. 
As the number of agents increases, predictability is reduced and
beyond a critical number of agents, the market becomes 
unpredictable (to agents). At the same time also the 
volatility of the market decreases. Beyond the critical number,
however, the market remains unpredictable but volatility starts
increasing. In addition excess volatility due to crowd effects arise, 
which manifest in
an anti-persistent behavior of the market \cite{Savit,CM}.

If the competitive aspects of the MG are related to predictability,
from the point of view of volatility the MG is similar in nature to
coordination games \cite{coordgames}. 
Indeed among all possible ways to exploit market's predictability, 
each agent has incentives to favor those which reduce
market's volatility. Since the MG is a closed dynamical system, 
fluctuations inevitably depend on the behavior of agents.

Here we study in detail the relation between volatility and the
behavior of agents. There are two key aspects of agent behavior
which have been found to dramatically affect volatility in the MG:
\begin{enumerate}
\item The market impact and the extent to which agents account for it. 
In the original MG, agents neglect their market impact and behave
as price takers\footnote{Price taking behavior means that, when 
an agent considers counter-factually the results of the various choice 
he/she could have made, 
he/she does not account for the fact that the market ``price'' depends on the
choice he/she actually made. 
If he/she had taken a different decision the ``price'' also would have 
changed. In few words, one may say that agents, rather than playing against
other rational agents, thinks as if they were playing against 
an exogenous price process.}. 
This behavior minimizes market's predictability but
this leads to large volatility increases (crowd effects) 
when the market is crowded. This excess volatility is dramatically reduced
as soon as agents start accounting, even approximately, for their
impact. When agents behave strategically, properly accounting for their
impact on the market, volatility is reduced to zero.
\item Agents reactivity, i.e. how readily and strongly do agents react to 
the outcome of the game. Fast learning rates may dramatically enhance
volatility in the MG, with respect to that with small learning rates.
\end{enumerate}

These results were found recently for the MG in its full complexity 
\cite{CMZ,MCZ} and for a modified version of it 
\cite{Cavagna2}\footnote{\ccite{comment} showed that the model introduced
by \ccite{Cavagna2} has exactly the same collective behavior of the 
original MG in the limit $N\to\infty$.}.
Here we derive these result in the simplest
case where there is no public information. This case allows
for a quite simple treatment where the result is not obscured by the
technicalities needed to deal with the MG with information. 

In addition it allows to understand the origin of enhanced volatility
for fast learning rates. This arises as a transition from a smooth 
dynamics to a more complex dynamical behavior which is reminiscent
of chaotic or turbulent systems.

Then we move to the MG with information and 
briefly review the results of \ccite{MCZ}. In addition we derive 
a new result for the critical learning rate, as a function of 
$\alpha$ above which crowd effects arise in the MG.

We close with some discussion of the results.

\section{The model without information}

The MG is a toy model of a market under the simplifying assumption that
{\em the market is a repeated game of $N$ interacting agents}. 
In real markets, agents can decide not to participate. In addition
real agents trade at different frequencies. In the MG all agents have to trade
at each time step. If $a_i(t)$ represents the investment of agent $i$ 
at time $t$, his/her profit from this transaction is
\be
u_i(t)=a_i(t)R[A(t)/N],~~~~~~A(t)=\sum_{j=1}^N a_j(t).
\label{marketint}
\ee
The return $R(A/N)$ depends on the aggregate investment $A(t)$. This
functional dependence embodies the market mechanism and it must be 
such that agents cannot extract a positive profit from just trading.
Hence the game is a negative sum game:
\be
\sum_{i=1}^N u_i(t)=A(t)R[A(t)/N]\le 0,~~~~ \forall t.
\ee

In addition we further assume that $a_i(t)$ can take only two values 
$\pm 1$. Hence agents choosing the minority sign $a_i=-\sign A$ gain a 
profit $|R(A/N)|$ whereas the majority of agents loose an amount 
$-|R(A/N)|$. The simplest 
choice $R(x)=-x$ will be made henceforth\footnote{This is justified 
{\em a posteriori} by power expansion of $R(x)$ in $x$ because $A(t)$ is
typically of order $\sqrt{N}$. The original choice in \cite{CZ1,Savit}
had $R(x)={\rm sign}\, x$ which can also be dealt with similarly
\cite{CMZh}.}

Naively, each player faces the problem of anticipating what the majority
will do, in order to do the opposite. This is similar in essence to the 
El Farol bar problem \cite{Arthur} 
and to many other situations where players have to 
share a limited resource.

The last ingredient of the MG is public information. 
Since the effects which we wish to focus on emerge even in the
absence of public information, we shall introduce this ingredient 
later on.

Let us generalize the model and allow agents to behave in a stochastic
way. Hence we introduce the probabilities $\pi_i$ with 
which agent $i$ plays action $a_i=+1$ (with ${\rm Prob}\{a_i=-1\}=1-\pi_i$).
With $\pi_i=0$ or $1$ we recover the deterministic case, 
whereas $\pi_i=1/2$ means that agent $i$ relies on the toss of a fair
coin, to decide his/her action. 
Expected values over $\pi_i$ are denoted by $\avg{\ldots}$. 

For any state $\vec\pi=(\pi_1,\ldots,\pi_N)$, we define
\be
H(\vec\pi)=\avg{A}^2,~~~~~~~\avg{A}=\sum_{i=1}^N(2\pi_i-1)
\label{H}
\ee
and 
\be
\sigma^2(\vec\pi)\equiv\avg{A^2}=H+4\sum_{i=1}^N\pi_i(1-\pi_i).
\label{sigma2}
\ee

$H$ measures information efficiency. Indeed $H>0$ implies
$\avg{A}\neq 0$ which means that one can predict the minority
action $-{\rm sign}\,\avg{A}$. When $H=0$ the outcome $A$ is
unpredictable. In both cases, $\sigma^2$ is simply related to the total loss
of agents $\sigma^2=-N\sum_i\avg{u_i}$ and hence it measures 
optimality. 
Volatility can be defined as the excess fluctuations
$\sigma^2-H$ of the market. 

\section{Nash equilibria}

The key concept to analyze the outcome of strategic interactions are
the Nash equilibria of game theory \cite{coordgames}. The MG is an
$n-$players game, where each agent $i$ may play pure strategies $a_i=+1$ or 
$-1$ -- or mixed strategies $\pi_i$ -- and the payoff matrix is specified 
by Eq. (\ref{marketint}). Nash equilibria \cite{coordgames}
are defined as those {\em states} $\vec \pi$ in which agents have no
payoff incentive to change their behavior. 

The purpose of this section is to provide a reference framework for our
later discussion on adaptive agents. 
More precisely it is of interest to undestand under what conditions adaptive 
learning can lead to the outcome one would expect from rational players,
i.e. to a Nash equilibrium.
Hence we shall limit our discussion to a rather elementary level\footnote{In 
what follows, Nash equilibria are vectors $\vec\pi$ such that no agent can 
increase his/her payoff by changing his/her mixes strategy $\pi_i$, if others
stick to theirs.}.

Let us consider a state $\vec\pi^{(j)}$ 
where $N-2j$ agents play mixed strategies $\pi_i=1/2$ and $2j$ 
play pure strategies. The {\em frozen} agents playing pure strategies are 
divided into two groups of $j$ agents playing opposite actions. 
{\em This is a Nash equilibrium}. 

In order to show this, we observe that the $2j$ agents playing pure
strategies have no incentive
to change it, because if they deviate from their strategy
they suffer a loss. Agents playing mixed strategies must, on the
other hand be indifferent between the two action, which occurs
if $\avg{A-a_i}=\sum_{j\ne i}(2\pi_j-1)=0$. 

Overall, by simple combinatorics we find that there are
\be
\Omega(N)=\sum_{j=0}^{[N/2]}2{N\choose N-2j}{2j\choose j}-1
\label{NashN}
\ee
Nash equilibria of this form (here $[N/2]$ is the integer part of $[N/2]$). 
For $N$ even, the information efficiency and optimality of
these equilibria are quantified by
\be
H(\vec\pi^{(j)})=0,~~~~
\sigma^2(\vec\pi^{(j)})=N-2j.
\ee

This makes it clear that agents playing mixed strategies $\pi_i^{(j)}=1/2$ 
are those who contribute to the value of $\sigma^2$. Hence they are 
responsible for inefficiencies (excess volatility). As the number $j$
of {\em frozen} agent pairs increases the equilibrium becomes more and more
efficient\footnote{In game theoretical terms one may say that for any Nash
equilibrium $\pi^{(j)}$ there is an equilibrium $\pi^{(j+1)}$ which
is Pareto dominant: In the latter each agent is not worse off than
in the first and there is at least one agent which is strictly better off.}
and $\sigma^2$ decreases. The equilibria $\vec\pi^{([N/2])}$, where
agents attain the minimal value of $\sigma^2$, shall be called {\em optimal}
henceforth. At the other extreme, the {\em symmetric} Nash equilibrium 
in which all agents play mixed strategies ($\pi_i^{(0)}=1/2$ for all $i$) 
is the most inefficient.

The number of frozen agents can increase if we allow for 
coordination devices such as binding agreements or public information. 
If agents $i$ and $k$, playing $(\pi_i,\pi_k)=(1/2,1/2)$ coordinate
their actions choosing to play either $(\pi_i,\pi_k)=(0,1)$ or 
$(\pi_i,\pi_k)=(1,0)$ with some probability, their payoffs increase
(which increases the payoffs of other agents as well). 
By this, however, we are assuming that the two agents play in a
correlated way\footnote{We refer the reader to \cite{coordgames} for a 
detailed discussion of coordination games and correlated equilibria.}. 
Iteration of this argument suggests that 
agents will naturally choose one of the optimal equilibria 
$\vec\pi^{([N/2])}$ if they can correlate their play. 

Summarizing, the single stage game has an huge number of Nash equilibria 
for $N\gg 1$ (Eq. \ref{NashN}). This raises the problem of equilibrium 
selection: Which of these equilibria describes the behavior of
deductive rational agents? 
Unfortunately game theory, on the basis of rationality and common knowledge 
alone, does not provide an unambiguous answer to this question.

Since Nash equilbria are very different in nature, with values of 
$\sigma^2$ ranging from $0$ to $N$, this means that, in fact, we cannot
draw any conclusion at all. Hence, as suggested by \ccite{Arthur}, 
one has to abandon deductive rationality and replace game-theoretic players 
by inductively rational adaptive agents.

\section{Adaptive agents}

Agents are adaptive, in the sense that they learn from past experience
which choice is the best one. The past experience of agent $i$ is stored in 
$\Delta_i(t)$ which is related to the cumulated payoff of the two possible
actions. $\Delta_i(t)>0$ means that the action $a_i=+1$ is (perceived as) 
more successful than $a_i=-1$ and {\em vice-versa}.
On the basis of their experience $\Delta_i(t)$ up to time $t$, the behavior
of agents is described by
\be
{\rm Prob}\{a_i(t)=\pm 1\}\equiv\frac{1\pm m_i(t)}{2},~~~
\hbox{ with }~~~m_i(t)=\chi_i[\Delta_i(t)]
\label{deca}
\ee
The function $\chi_i(x)$ is an increasing function of $x$ with 
$\lim_{x\to\pm\infty}\chi_i(x)=\pm 1$.
Hence the probability with which agent $i$ plays action $a_i=\pm 1$
increases with $\Delta_i(t)$.

The connection with mixed strategies is $m_i=2\pi_i-1$. Here we choose a
different notation to stress the fact that $m_i(t)$ is not a mixed strategy,
but rather it encodes a behavioral rule. Note that $m_i=\avg{a_i}$.

The way in which $\Delta_i(t)$ is updated is the last element of the
learning dynamics. We assume that
\be
\Delta_i(t+1)=\Delta_i(t)-\frac{1}{N}[A(t)-\eta_i a_i(t)]
\label{learn}
\ee
The idea is: if agent $i$ observes $A(t)<0$ he/she will generally
increase $\Delta_i$ and hence his/her probability of playing $a_i=+1$
at the next time step. The $\eta_i$ term above describe the fact that
agent $i$ may account for his/her own contribution to $A(t)$. 
For $\eta_i=1$ indeed, agent $i$ considers only the behavior of
other agents $A(t)-a_i(t)$ and does not react to his/her own
action $a_i(t)$.

The question of interest is: {\em Do agents converge to a Nash 
equilibrium under this dynamics?} If yes, {\em to which 
equilibrium do agents converge?}

\subsection{Naive agents $\eta_i=0$}

In the MG agents are naive and do not account for their impact on
the aggregate, i.e. $\eta_i=0$. 

Let us focus on the long time dynamics for $N\gg 1$. In particular we
follow \cite{MCZ} and introduce a time variable $\tau=t/N$. A 
superscript ${}^{(c)}$ denotes variables in the continuum
time (e.g. $\Delta_i^{(c)}(\tau)=\Delta_i(N\tau)$). 
Then 
\be
\Delta_i^{(c)}(\tau+d\tau)=
\Delta_i^{(c)}(\tau)-\frac{1}{N}\sum_{t=N\tau}^{N(\tau+d\tau)}A(t).
\label{upddc}
\ee
For $N\gg 1$ we can use the law of large numbers and conclude that
the last term is well approximated by $-d\tau\avg{A}_\tau$. Here
the subscript $\tau$ implies that the average is taken over the 
probabilities $[1\pm m^{(c)}_i(\tau)]/2$ of the actions
$a_i=\pm 1$, and we are assuming that 
$\chi_i(x)$ is smooth enough (see later) so that these probabilities are
approximately constant over the time interval $[\tau,\tau+d\tau)$.
Hence we find that
\be
\dot{\Delta}_i^{(c)}(\tau)=-\avg{A}_\tau=-\sum_{i=1}^N
m_i^{(c)}(\tau).
\ee

It is quite easy to find that 
{\em $H$ is minimized along the trajectories of the 
learning dynamics with $\eta_i=0$ for all $i$.}

Indeed, with the notation $\avg{\ldots}_\tau$ for averages taken
at time $\tau$
\be
\dot{H}=2\avg{A}_\tau\sum_{i=1}^N
\frac{\partial \avg{A}_\tau}{\partial m_i^{(c)}}\dot{m}_i^{(c)}=
-2H\sum_{i=1}^N\chi_i'(\Delta_i^{(c)})
\ee
where we have assumed that $\avg{\chi_i(\Delta_i)}_\tau\simeq
\chi_i(\Delta_i^{(c)})$, which is correct to leading order in $1/N$.

Since $\chi_i'(x)\ge 0$ for all $x$ and $i$, we conclude that 
$\dot{H}\le 0$ i.e. {\em naive agents minimize the predictability 
$H$}, and $H\to 0$ as $t\to\infty$. There are many states with 
$H=0$ and the dynamics select that which is ``closer'' to the initial
condition. To be more precise, let $\Delta_i(0)$ be the initial 
condition (which encodes the prior beliefs of agents on which action
is best). Then, as $\tau\to\infty$, $\avg{A}_\tau\to 0$ and 
$\Delta_i^{(c)}(\tau)$ converges to
\be
\Delta_i^{(c)}(\infty)=\Delta_i(0)+\delta A,~~~ \hbox{with}~~~
\delta A=\int_0^\infty d\tau\avg{A}_{\tau}.
\ee
The condition $\avg{A}_\infty=0$ provides an equation for $\delta A$
\be
0=\sum_{i=1}^N\chi_i\left(\Delta_i(0)+\delta A\right).
\label{eqnaive}
\ee
By the monotonicity property of $\chi_i$, this equation has one and
only one solution.

The asymptotic state of this dynamics is information--efficient ($H=0$), 
but it is not optimal. Indeed, in general, this state {\em is not} a Nash
equilibrium. Typically we find $\sigma^2\propto N$. Only in the special case
$\Delta_i(0)=0$ and $\chi_i(0)=0$ for all $i$, we recover the
symmetric Nash equilibrium $\vec\pi^{(0)}=(1/2,\ldots,1/2)$ where 
$\sigma^2=N$.

\subsection{Less naive agents $\eta_i>0$}

It is easy to check that with $\eta_i>0$, following the same
steps as in the previous section, the learning dynamics of agents
minimize the function
\be
H_\eta=\avg{A}^2-\sum_{i=1}^N \eta_i m_i^2.
\label{Heta}
\ee

Because of the new term, $H_\eta$ attains its minima on the boundary
of the domain $[-1,1]^N$. In other words, $m_i=\pm 1$ for all $i$ which
means that agents play pure strategies $a_i=m_i$. 
The stable states are optimal Nash equilibria for $N$ even. 
By playing pure
strategies agents minimize the second term of $H_\eta$. Of all corner
states where $m_i^2=1$ for all $i$, agents select those with 
$\avg{A}=0$ by dividing into two equal groups
playing opposite actions. All these states have minimal ``energy''
$H_\eta=-\sum_i \eta_i$. Which of these states is selected 
depends on the initial conditions $\Delta_i(0)$. 

Note that
the set of stable states is {\em disconnected}. Each state has its
basin of attraction in the space of $\Delta_i(0)$. The stable
state changes discontinuously as $\Delta_i(0)$ is varied.
This contrasts with the case $\eta_i=0$ where Eq. \req{eqnaive} implies 
that the stationary state changes continuously with $\Delta_i(0)$
and the set of stationary states is connected.

For $N$ odd, similar conclusions can be found. 
This can be understood by adding a further agent to a state 
with $N-1$ (even) agents in a Nash equilibrium.
Then $H_\eta=(1-\eta_N)m_N^2$, so for $\eta_N<1$ the new agent
will play a mixed strategy $m_i=0$, whereas for $\eta_N>1$ it will 
play a pure strategy. In both cases other agents have no incentive
to change their position. In this case we find $\sigma^2\le 1$.

Summarizing, {\em when agents account for their impact
on the aggregate, they attain not only an 
information--efficient state, but also a optimal Nash equilibrium with} 
$\sigma^2=H=0$. 

It is remarkable how the addition of the parameter $\eta_i$ 
radically changes the nature of the stationary state. 
Most strikingly, fluctuations are reduced by a factor $N$.

\subsection{Naive agents learning at a fast rate}

The difference between the behavior of naive and non naive agents
becomes even stronger
if agents are naive ($\eta_i=0$) and very reactive. In order to
quantify more precisely what we mean, let us assume that\footnote{In 
this section we assume that agents follow a Logit model of discrete choice
where the probability of choice $a$ is proportional to the
exponential of the ``score'' $U_a$ of that choice: $\pi(a)\propto 
e^{\Gamma U_a/2}$. With only two choices $a=\pm 1$, 
$\pi(a)=(1+am)/2$ and $\Delta=U_+-U_-$, we recover Eq. \req{logit}.
This learning model has been introduced
by \ccite{Cavagna2} in the context of the MG.}
\be
\chi_i(\Delta)=\tanh (\Gamma\Delta), ~~~~\forall i.
\label{logit}
\ee
Here $\Gamma$ is the learning rate, which measures the scale of 
the reaction in agent's behavior (i.e. in $m_i$) to a change
in $\Delta_i$.
We also assume that agents have no prior beliefs: $\Delta_i(0)=0$.
Hence $\Delta_i(t)\equiv y(t)/\Gamma$ is the same for all agents.
From the results discussed above, we expect, in this case the 
system to converge to the symmetric Nash equilibrium $m_i=0$ 
for all $i$. This is not going to be true if agents are too reactive,
i.e. if $\Gamma>\Gamma_c$, as we shall see shortly.

\begin{figure}
\centerline{\psfig{file=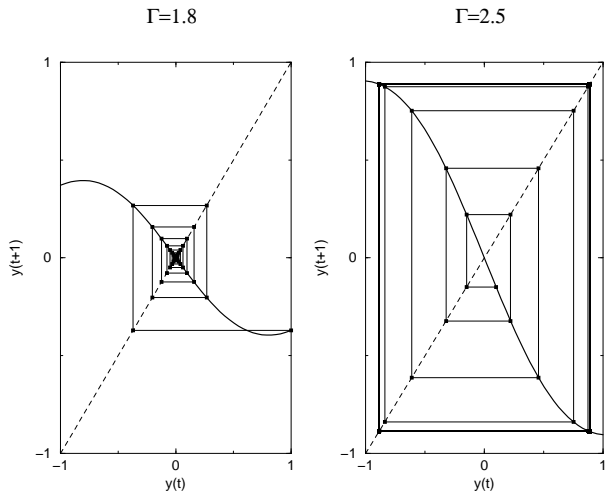,width=8cm,angle=270}}
  \caption{Graphical iteration of the map $y(t)$ for $\Gamma=1.8<\Gamma_c$
and $\Gamma=2.5>\Gamma_c$}
  \label{figmap}
\end{figure}

Indeed $y(t)=\Gamma\Delta_i(t)$ satisfies the equation
\bea
y(t+1)&=& y(t)-\frac{\Gamma}{N}\sum_{i=1}^N a_i(t)\nonumber\\
&\simeq & y(t)-\Gamma\tanh[y(t)]
\label{dyn0}
\eea
where the approximation in the last equation relies on the law
of large numbers for $N\gg 1$. Eq. \req{dyn0} is a dynamical system. 
The point $y^0=0$ is stationary,
but it is easy to see that it is only stable for $\Gamma<\Gamma_c=2$.
For $\Gamma>2$, a cycle of period $2$ arises, as shown in figure
\ref{figmap}. This has dramatic effects on the optimality of
the system. Indeed, let $\pm y^*$ be the two values taken by
$y(t)$ in this cycle\footnote{$\pm y^*$ are the two non-zero solutions of 
$2y=\Gamma\tanh(y)$.}. Since $y(t+1)=-y(t)=
\pm y^*$ we still have $\avg{A}=0$ and hence $H=0$.
On the other hand $\sigma^2=N^2 {y^*}^2$ is of order
$N^2$, which is even worse than the symmetric Nash equilibrium
where $\sigma^2=N$.

This transition from a state where $\sigma^2\propto N$
to a state with $\sigma^2\propto N^2$ is generic in the minority game, 
as first observed by \ccite{Savit}.
Much work has been done by Johnson {\em et al.}
to describe this effect quantitatively in terms of ``crowds'' and
``anti-crows''. This approach is however static and does not
reveal the dynamical nature of the transition, which was first
discussed in \cite{CM}. The transition from these
two regimes as the learning rate $\Gamma$ is changed has been
first observed in  \cite{Cavagna2} for the MG.

We shall see below that the simple approach followed here
can be generalized to the full minority game and it allows to 
derive the critical learning rate $\Gamma_c(\alpha)$ as a function
of the parameter $\alpha$ of the MG.

\section{The MG with information}

The same qualitative behavior occurs when agents have access to a
public information represented by an integer variable $\mu$ ranging
from $1$ to $P$. $\mu$ may either be related to past
outcomes of the game \cite{CZ1} or, as first suggested by \ccite{Cavagna},
it may be just randomly and uniformly drawn from $\{1,\ldots,P\}$.
The idea is that, agents resort to 
simple schemata or rules, which prescribe an action for each value
of $\mu$. Each agent is initially
endowed with $S$ such rules which are drawn at random among all
possible $2^P$ binary functions. Agents take their choice on the basis
of ``scores'' $U_{s,i}$ which they assign to each strategy $s$, and they
update scores in a way which is similar to Eq. \req{learn}.
\ccite{Savit} has shown that the relevant variable is $\alpha=P/N$: 
Intensive (i.e. $N$ independent) quantities, such as $\sigma^2/N$,
display a behavior which does not depend on $P$ and $N$ separately,
but only on their ratio $\alpha$. 
We refer the interested reader to  \cite{MCZ} for a discussion
of the general case. Here we focus on the $S=2$ case, where following
 \cite{CM,CMZ}, we let $s=\pm 1$ be the label of the two possible
rules of each agent and $s_i(t)$ be the choice actually taken
by agent $i$ at time $t$. Following the notation of \cite{CM,CMZ}, 
the action taken by agent $i$ at time $t$, if
he/she chooses to follow strategy $s_i(t)$ is $a_i(t)=
\omega^{\mu(t)}_i+s_i(t)\xi_i^{\mu(t)}$, where $\mu(t)$ is the value
taken by public information\footnote{$\omega_i^\mu$ and $\xi_i^mu$ are
such that $\omega_i^\mu\pm\xi_i^mu$ are two randomly choosen boolean
functions of $\mu$ taking values in $\{\pm 1\}$.}.
The decision process of agent $i$ is 
hence encoded, as in Eq. \req{deca} above, in the equation
\be
{\rm Prob}\{s_i(t)=\pm 1\}\equiv\frac{1\pm m_i(t)}{2},~~
\hbox{ with }~~m_i(t)=\chi_i(\Delta_i).
\label{decs}
\ee
Here $\Delta_i(t)$ is the difference between the scores of the two
schemata, which are updated according to the analog of Eq. \req{learn}:
\be
\Delta_i(t+1)=\Delta_i(t)-\frac{A(t)\xi_i^{\mu(t)}-\eta_i{\xi_i^{\mu(t)}}^2
s_i(t)}{N}.
\label{updDMG}
\ee

Refs. \cite{MCZ,comment} have shown that typical relaxation times in a system 
of $N$ agents are of order $N$. In order to have a meaningful dynamics in the
limit $N\to\infty$ it is necessary to introduce a rescaled 
time $\tau=t/N$. 
This leads to a deterministic continuum time dynamics \cite{CM,CMZ} which 
is independent of $N$. The validity of this description relies on the fact 
that typically $A(t)\sim\sqrt{N}$ and hence 
$|\Delta_i(t+1)-\Delta_i(t)|\sim 1/\sqrt{N}$ is vanishingly small, which 
makes the continuum time approach exact as $N\to\infty$. As we shall see, 
this fails to be true if $\alpha<\alpha_c$ and agents are too reactive, since 
then $A(t)\sim N$.

\begin{figure}
\centerline{\psfig{file=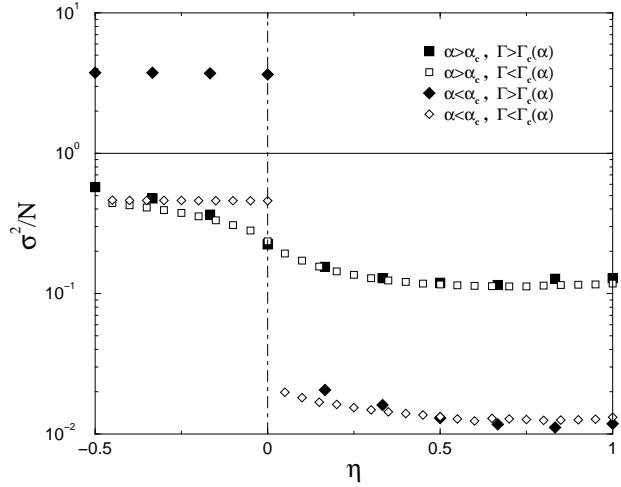,width=8cm}}
\caption{$\sigma^2/N$ as a function of $\eta$ for $S=2$ and $N=101$.
Diamonds refer to $P=8$ ($\alpha\simeq 0.079<\alpha_c$) with
$\Gamma=\infty$ (open symbol) and $\Gamma\ll \Gamma_c(\alpha)$
(full symbol). Squares refer to $P=64$ ($\alpha\simeq 0.63>\alpha_c$)
with $\Gamma=\infty$ (open symbol) and $\Gamma\ll \Gamma_c(\alpha)$
(full symbol).}
\label{figsigvseta}
\end{figure}

\begin{figure}
\centerline{\psfig{file=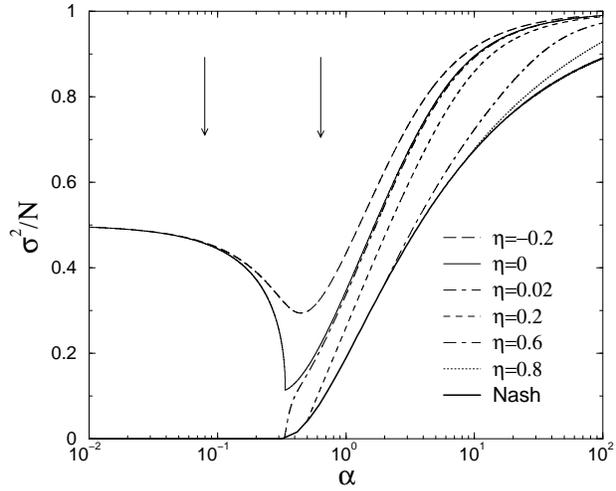,width=8cm}}
\caption{Approximate theoretical estimate of global efficiency 
$\sigma^2/N$ as a function 
of $\alpha$ for $S=2$ and several values of $\eta$ within the {\em replica 
symmetric ansatz}. The arrows mark the values of $\alpha$ to which 
Fig. 1 refers.}
\label{figeta_RS}
\end{figure}

Within the continuum time description one finds \cite{CMZ,MCZ} that
with $\eta_i=0$ for all $i$, the behavior of agents again minimizes 
predictability $H$.
It can be shown \cite{MCZ} that the payoff of each agent $i$ increases 
with $\eta_i$ in the range $\eta_i\in [0,1]$. In addition, also global
efficiency increases. If $P>\alpha_c N$, where $\alpha_c=0.3374\ldots$,
the improvement is smooth as a function of $\eta$. If, on the
other hand, $P<\alpha_c N$ there is a sudden jump in $\sigma^2$ 
as soon as agents switch on a small parameter $\eta_i=\eta$. 
This is shown clearly in figure \ref{figsigvseta}. This feature is
captured qualitatively by the analytic calculation (in the replica
symmetric {\em ansatz}) of \cite{CMZ,MCZ}, whose result is
shown in Figure \ref{figeta_RS}. For $\alpha=P/N<\alpha_c$, as the 
line $\eta=0$ is crossed from below, the system undergoes a first
order phase transition with a discontinuous jump in $\sigma^2$. 
Note that for $\Gamma\gg\Gamma_c$ the jump in $\sigma^2/N$ is
of more than two decades!
This discontinuity arises because the nature of the stationary state 
changes 
abruptly: For $\eta\le 0$ a finite fraction of agents play mixed
strategies ($m_i^2<1$) whereas for $\eta>0$ most of them play
only pure strategies. For $\eta=1$ all agents play pure strategies
and indeed the system converges to a Nash equilibrium\footnote{Here
Nash equilibria are defined with respect to the set of pure strategies
$s_i=\pm 1$ which agents can take. See \ccite{MCZ} for a detailed 
discussion.}. 

The transition for $\alpha>\alpha_c$ occurs at a value $\eta_c(\alpha)>0$
and it is smooth (second order). A deeper discussion on the nature
of this transition and on its consequences may be found in  
\cite{MCZ,andemar}.

\section{Learning rate in the MG for $\alpha<\alpha_c$}

\begin{figure}
\centerline{\psfig{file=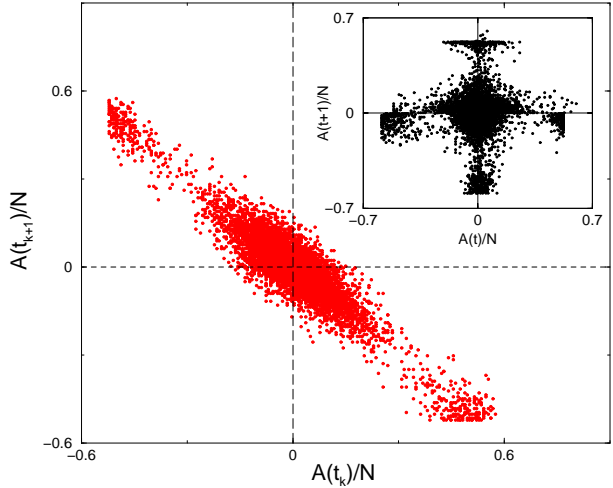,width=8cm,angle=270}}
\caption{Plot of $A(t_{k+1})/N$ vs $A(t_k)/N$ for the MG with 
$N=301$ agents, $\Gamma=\infty$ and $P=16$ ($\alpha=0.053\ldots$).
The sequence $t_k$ is such that the information is the same
($\mu(t_{k+1})=\mu(t_k)$). In the inset, the same plot with real
time $t$.}
\label{figscatt}
\end{figure}

As before, the performance of naive agents ($\eta_i=0$ $\forall i$)
may be much worse for $\alpha<\alpha_c$ if they are too reactive. 
This is shown numerically
in figure \ref{figsigvseta}. The effect is exactly the same as that
discussed previously, in the absence of information ($P=1$ or 
$\alpha\simeq 0$):
As the learning rate $\Gamma$ increases, the stationary solution
$\Delta_i^*$ looses its stability and a bifurcation to a complex dynamics 
occurs. This is only possible in the low $\alpha$ phase, where the stationary 
state is degenerate and the system can attain $\avg{A^\mu}=0$ (intended as 
a time average) 
by hopping between different states\footnote{In the asymmetric
phase $\alpha>\alpha_c$ the stationary state is unique and this
effect is not possible.}. The plot of $A(t+1)/N$ vs 
$A(t)/N$, in figure \ref{figscatt} shows that indeed wild fluctuations
occur in one time step: a finite fraction of agents change their
mind at each time step. 
This is what causes, for fixed $P$, the cross over, first
observed in \cite{Savit}, from the linear regime 
$\sigma^2\sim N$ to a quadratic dependence $\sigma^2\sim N^2$.

Clearly the continuum time limit, on which our analysis rests, 
breaks down because $A(t)\sim N$. 
Still one can compute the critical learning rate 
$\Gamma_c(\alpha)$
which marks the onset of complex dynamics. Let us focus attention
on one value of $\mu=1$ and on the learning model of Eq. \req{logit}.
Let us define the sequence of times $t_k$ such that
$\mu(t_k)=1$ for the $k^{\rm th}$ time. Fig. \ref{figscatt} shows that 
the dynamics of $A$ in this modified time is not smooth.
We define 
$y_i(k)=\Gamma\Delta_i(t_k)$, which satisfies
\be
y_i(k+1)=y_i(k)-\frac{\Gamma}{N}\sum_{t=t_k}^{t_{k+1}-1}A(t)\xi_i^{\mu(t)}
\label{eqy}
\ee

\begin{figure}
\centerline{\psfig{file=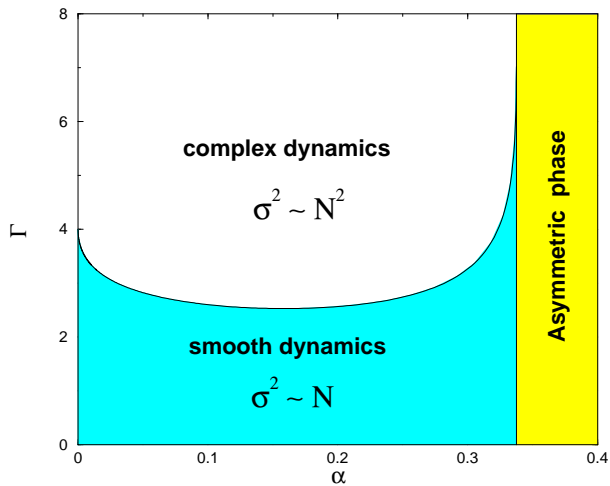,width=8cm,angle=270}}
\caption{Phase diagram of the MG ($\eta_i=0$) in the 
$(\alpha,\Gamma)$ plane.}
\label{figphaseG}
\end{figure}

When $N\gg 1$, the sum involves $\sim P=\alpha N\gg 1$ terms and
we may estimate it by the law of large numbers. Let $y_i^*$ be the stationary
solution ($y_i(k+1)=y_i(k)=y_i^*$) 
of Eq. \req{eqy}, then we can set $y_i(k)=y_i^*+\delta y_i(k)$
and study the linear stability of this solution.
With the notation $\ovl{R}=\sum_\mu R^\mu/P$, we find
\be
\delta y_i(k+1)=\sum_{j=1}^N T_{i,j}\delta y_j(k), ~~~
T_{i,j}=\delta_{i,j}-\alpha\Gamma\ovl{\xi_i\xi_j}(1-m_j^2)
\ee
where $m_j=\tanh(y_j^*)$. The solution $y^*_i$ is stable if the
eigenvalues of $T_{i,j}$ are all smaller than $1$ in absolute value.
As $\Gamma$ increases, the smallest eigenvalue of 
$T_{i,j}$ becomes smaller that $-1$. Thanks to the results of
\ccite{Mitra}, we have an analytic expression for this
eigenvalue, which is $\lambda_+=1-\Gamma(1+\sqrt{\alpha})^2(1-Q)/2$.
The stability condition $\lambda_+>-1$ then turns into 
\be
\Gamma<\Gamma_c(\alpha)\equiv \frac{4}{[1-Q(\alpha)](1+\sqrt{\alpha})^2},
~~~~Q(\alpha)=\frac{1}{N}\sum_{i=1}^N m_i^2
\label{betac}
\ee
which is our desired result. The function $Q(\alpha)$ is known exactly
from the analytic solution\footnote{$Q(\alpha)$ is given parametrically by
$Q=1-e^{-z^2}/(\sqrt{\pi}z)-(z^2-1/2){\rm erf}(z)/z^2$ and
$\alpha={\rm erf}^2(z)/[2z^2(1+Q)]$.} \cite{CMZ,MCZ}. This yields
a phase diagram in the $(\alpha,\Gamma)$ plane which is shown in 
figure \ref{figphaseG}. For $\alpha\to 0$ we find 
$\Gamma_c\to 4$\footnote{This differs from our previous result 
$\Gamma_c=2$ without information, because with $P=1$ in the MG
half of the population has $a_{+,i}=a_{-,i}$ two equal 
strategies. This reduces by a factor $2$ the effective number of adaptive
agents, and accordingly $\Gamma_c$ takes a factor $2$.}.
As $\alpha\to\alpha_c$, $\Gamma_c$ converges to a finite value
($\simeq 7.0273\ldots$) with infinite slope. 
Numerical simulations suggests that, rather than a sharp
transition, at $\Gamma_c(\alpha)$ the system undergoes a crossover 
between two distinct dynamical regimes. In the asymmetric phase 
($\alpha>\alpha_c$) the dynamics is always smooth, hence
$\Gamma_c(\alpha)=\infty$. 

The analysis of the stationary state for $\Gamma\gg \Gamma_c$ and the 
calculation of global quantities such as $\sigma^2$ is much more 
difficult than that in the smooth dynamical phase ($\Gamma\ll \Gamma_c$).
\ccite{Johnson99} devised an approximate scheme, which neglects 
the dynamical aspect of the problem, but gives expressions for 
$\sigma^2$ in good agreement with numerical data for $\alpha<\alpha_c$
and $\Gamma\gg \Gamma_c(\alpha)$. A microscopic, systematic derivation 
of $\sigma^2$ is a challenging problem, and we believe the dynamical
system (\ref{eqy}) is the starting point.

\section{Conclusion}

In its simplicity, the MG captures and reproduces a great deal of
mechanisms, aspects and properties of real markets. In particular
it provides a deep understanding of the relation between behavioral
assumptions at the micro level and global macro behavior in a toy market.
Here we have shown what properties of the behavior of agents are 
responsible for excess volatility in such toy markets. These are
price taking behavior and excessive reactivity (fast learning).

This leaves us with the question of whether and to what extent these 
conclusions extend to real market contexts. 
Experimental studies in this direction promise to be very 
illuminating.

\label{lastpage}

\end{document}